# The Formation of Dwarf Galaxies in Tidal Debris: A Study of the Compact Group Environment

Sally D. Hunsberger and Jane C. Charlton[1]

*Astronomy and Astrophysics Department*
*Pennsylvania State University*
*University Park, PA 16802*
*sdh@astro.psu.edu*
*charlton@astro.psu.edu*

and

Dennis Zaritsky

*UCO/Lick Observatories & Board of Studies in Astronomy and Astrophysics,*
*University of California, Santa Cruz, CA 95064*
*dennis@ucolick.org*

## ABSTRACT

From R-band images of 42 Hickson compact groups, we present a sample of 47 candidate dwarf galaxies that are associated with the tidal tails and arms in the groups. The candidates, found in 15 tidal features, have R magnitudes and masses (for M/L = 1) in the ranges $-16.5 \leq M_R - 5 \log h_{75} \leq -11.5$ and $2 \times 10^6 M_\odot \leq M \leq 2 \times 10^8 M_\odot$, respectively. Their masses and locations are compared to the predictions of theoretical/N-body tidal dwarf formation scenarios. Considering the longevity of tidal debris in the compact group environment and the results of this survey, we estimate the contribution of the tidal dwarf formation mechanism to the population of dwarf galaxies observed at large in compact groups. If the majority of our dwarf galaxy candidates are confirmed as being gravitationally bound stellar systems, then a significant fraction, perhaps as much as one-half, of the dwarf population in compact groups is the product of interactions among giant parent galaxies.

*Subject headings:* galaxies — clustering, formation, interactions

---

[1] Center for Gravitational Physics and Geometry, Pennsylvania State University

1. **Introduction**

There is anecdotal observational evidence that dwarf galaxies form in the tidal debris from giant galaxy interactions. Regions of active star formation are observed at the end of tidal tails in the Superantennae (IRAS 19254-7245) (Mirabel, Lutz, & Maza 1991) and Antennae (NGC 4038/39) (Mirabel, Dottori, & Lutz 1992) systems. In Arp 105, a blue compact dwarf and an Im galaxy have formed at the ends of two tidal tails (Duc & Mirabel 1994). Finally, numerous large HI clouds are observed in the interacting pair IC 2163/NGC 2207 (Elmegreen et al. 1995) and Elmegreen, Kaufman, & Thomasson (1993) suggest that some of these may form stars, detach, and become separate dwarf galaxies.

The idea that self-gravitating objects in tidal tails could evolve to become dwarf galaxies was originated by Zwicky (1956). Recent numerical simulations confirm that clumps can form within tidal tails, but the properties of the simulated dwarfs depend on the details of the modeling. Barnes and Hernquist's N-body/SPH simulations (1992) produce dwarf galaxies with masses in the range $1 \times 10^7 M_\odot$ to $4 \times 10^8 M_\odot$ during a single encounter between two giant galaxies. In their model, high mass dwarfs form anywhere along the tidal tail, but small mass dwarfs form only near the end of the tail where tidal disruption is less severe. Elmegreen et al. (1993) apply a two-dimensional N-body code (Thomasson 1989) to the problem of condensation of HI clouds in a tidal tail created from an extended gaseous disk, treating the gas as a dissipational particle component. They find that the velocity dispersion in the gas defines the mass of the clouds produced. Because the details of the interaction set a "characteristic" velocity dispersion, a smaller range of dwarf galaxy masses (within a factor of 5) is produced in a single encounter. Lastly, their simulations show that a large pool of gas ($\sim 10^9 M_\odot$) develops at the tip of the tidal tail from material drawn from the outer edge of the parent disk.

Hickson compact groups (HCGs) present a unique environment in which to study the tidal dwarf formation mechanism. Hickson (1982) cataloged 100 compact groups of galaxies from a systematic search of the Palomar survey prints, selecting them on the basis of population, isolation, and compactness. Radial velocity measurements (Hickson et al. 1992) and morphological studies (Mendes de Oliveira & Hickson 1994) suggest that most of the Hickson compact groups are physical associations, however Mamon (1990) has argued that the data are consistent with many HCGs being superpositions of binary-rich loose groups and Hernquist, Katz, & Weinberg (1995) contend that compact groups are chance projections of large filamentary structures. If the groups are real, then they have space densities comparable to the centers of galaxy clusters, but velocity dispersions more like those in loose groups. This combination is conducive to the formation of tidal tails, although the tails probably last less than a Gyr (Barnes 1993). We only utilize HCGs as likely sites for tidal features. Our search and detection of candidate dwarf galaxies is otherwise independent of the nature and lifetime of HCGs.

From our sample of 42 Hickson compact groups, we focus on the seven groups in which we have detected tidal features, and identify dwarf galaxy candidates within the tidal debris. The observational procedures and analysis are described in §2. The properties of the dwarf candidates, i.e., the luminosity function, derived masses, and the relationship of dwarf luminosity to the distance from the parent galaxy, are compared to model predictions in §3. In §4 we estimate the number of dwarfs formed in tidal debris over the lifetime of a compact group, and compare that number to the total number of dwarf galaxies predicted to be in this environment by the observed galaxy luminosity function.

2. **Data Acquisition and Analysis**

We selected compact groups from the Hickson catalog (1982) on the basis of their angular size and galaxy apparent magnitudes (Hickson 1982; Hickson et al. 1992). No other criteria which might bias our search for tidal features (e.g., color, richness, or previous observations) were used. To match the angular sizes of the groups to the field-of-view, we selected groups with angular diameters $< 7'$. In redshift, we chose groups with $z \leq 0.05$ so that faint dwarfs ($M_R > -15$) would be above our detection limit. For a typical redshift, $z = .03$ or $cz = 9000$ km s$^{-1}$, our limiting apparent magnitude of 21.5 corresponds to $M_R = -13.9$ (using $H_0 = 75$ km s$^{-1}$ Mpc$^{-1}$, $q_0 = 0.1$). Of the 66 Hickson compact groups that satisfy these criteria, 49 were observable in November from Palomar.

We obtained Johnson R-band images of 42 Hickson compact groups (listed in Table 1) using the 1.5-m telescope at Palomar during November 16-20, 1993.

TABLE 1
Observed Compact Groups

| HCG No. | | |
|---|---|---|
| 001 | 028 | 052 |
| 003 | 030 | 054 |
| 004 | 031 | 056 |
| 005 | 032 | 057 |
| 006 | 033 | 059 |
| 007 | 034 | 089 |
| 012 | 037 | 092 |
| 013 | 038 | 094 |
| 014 | 040 | 095 |
| 016 | 043 | 096 |
| 020 | 046 | 097 |
| 024 | 047 | 098 |
| 025 | 049 | 099 |
| 026 | 051 | 100 |

A thinned Tektronix 2048 × 2048 CCD was binned by 2 in both directions and provided a 12.5′ × 12.5′ field-of-view with 0.73 arcsec/pixel. Two 15-minute exposures were taken of each group over the first four nights. Calibration frames (5-minute exposures of each compact group) and standard star fields (Landolt 1992) were obtained during the last night under photometric conditions.

The data were reduced using IRAF[2]. A median of 44 bias frames was subtracted from each image. The median of 55 images, including standard star fields, calibration frames, and offset frames (taken 10′ from the group center), was used to flat field images because it produced a flatter background than either twilight or dome flats. Next, we removed bad pixels and cosmic rays, combined the two 15-minute images of each group, and calibrated the images using the 5-minute photometric exposures and standard stars observed at a range of airmasses.

We visually examine the images to identify the groups with interacting galaxies. About one-half of the observed groups contain interacting galaxies. The interactions are characterized by disrupted galaxies, extended luminous halos, tidal bridges, tidal arms, and/or tidal tails. We define a "tidal bridge" as an extension of material between the interacting galaxies, a "tidal arm" as an elongated or distorted spiral arm, and a "tidal tail" as a significant tail-like feature that is unassociated with a spiral arm. For the purpose of searching for tidal dwarfs, we select the seven groups which specifically exhibit tidal tails or tidal arms. Figures 1a through 1d are R-band images of these seven groups (Plates 1 through 4). Although the selection and classification of tidal features is somewhat subjective, we require that each pixel in the visually identified feature be at least $3\sigma$ above the background level. In fact, six of the groups have tidal features with pixel values $5\sigma$ above the local sky, and the true significance of any feature is much higher because it is the product of the number of pixels and the probability that each of them would be at least $3\sigma$ above background. This procedure places a lower limit on the number of tidal features in the sample, because features with surface brightnesses below 25.0 mag/arcsec$^2$ cannot be detected.

In Table 2 we compare our identification of tidal features to the results of a morphological study of HCGs by Mendes de Oliveira and Hickson (1994; hereafter MdOH) and summarize the characteristics of the seven groups in our sample in which we identify tidal features. In general, we confirm the MdOH classifications. Column 1 lists the Hickson catalog group number (Hickson 1982), column 2 lists the member identification (Hickson 1982) and morphological classification (Hickson, Kindl, & Auman 1989) of the interacting galaxies, column 3 gives the number of tidal dwarf candidates for each group, column 4 describes the type of tidal feature(s) and notes with which member it is associated, and column 5 presents the results from MdOH.

We use FOCAS (Faint Object Classification and Analysis System) (Jarvis & Tyson 1981) to identify non-stellar objects in the images. The software creates a catalog of objects on the basis of user-defined detection parameters. We require that objects have at least six contiguous pixels $1.5\sigma$ above the local background. We set these values interactively to include low surface brightness dwarf candidates that are apparent by visual inspection. FOCAS computes a significance statistic for each object which we use as a detection criterion. By examining detected objects and their significance, we define our lowest acceptable significance value to be 1.0. Those objects that are classified as galaxies (i.e., extended) and that lie within a tidal feature have been designated "dwarf candidates". The 47 candidate dwarfs in the seven groups with tidal features are marked in Figures 1a through 1d. From total galaxy counts provided by FOCAS, we estimate that an average of 0.7 background giants or dwarf members per group are coincident with the tidal features. Thus we expect $\sim 5$ of the tidal dwarfs to be unassociated with tidal features. Repeating the procedure with our star counts, we estimate that $\sim 8$ stars should be coincident with the tidal debris. FOCAS actually detects five stars in the tidal features, so contamination by stars is not a major problem.

---

[2]IRAF is distributed by the National Optical Astronomy Observatories, which are operated by the Association of Universities for Research in Astronomy, Inc. (AURA) under cooperative agreement with the National Science Foundation.

The magnitudes and positions of the tidal dwarfs will be the key to testing theoretical models for their formation. Because FOCAS consistently underestimates the local background within the debris, we use the IRAF POLYPHOT task to calculate the magnitudes of the dwarf candidates. Our standard star reductions indicate that there is a slight color term between the Landolt (1992) system and our observations. If the group tidal dwarfs are similar in color to the clumps in the Superantennae $((V-R) = 0.6$; Mirabel et al. 1991), our color term error is only 0.024 magnitudes. There is no color information available for our dwarf candidates, so we omit this apparently insignificant correction. Our standard star calibration uncertainties are $\sim .02$ magnitudes. The largest uncertainties come from the determination of the local background of surrounding debris, which we estimate to be typically $\sim 0.6$ mag by examining the statistics (mean and standard deviation) of the local background near the object's outermost significant isophote. The dwarf candidates have absolute magnitudes $-17.7 \leq M_R - 5\log h_{75} \leq -11.4$ with a median magnitude of $-14.2 + 5\log h_{75}$ (where $h_{75} = H_0/75$ km s$^{-1}$ Mpc$^{-1}$ and $q_0 = 0.1$). The corresponding luminosity function is given in Figure 2. Because there are 47 tidal dwarfs in the 42 compact groups in our sample, our estimate for the volume density of compact group tidal dwarfs is $1.8 \times 10^{-5} h_{75}^3$ Mpc$^{-3}$ (using $1.6 \times 10^{-5} h_{75}^3$ Mpc$^{-3}$ for the volume density of compact groups from Mendes de Oliveira & Hickson 1991). This estimate is used to normalize our luminosity function.

## 3. Properties of Tidal Dwarf Candidates

We fit a line to the tidal dwarf luminosity function (shown in Figure 2) in the magnitude range defined by the bright end of dwarf galaxies ($M_R = -16.5$) and our conservative estimate of the magnitude at which we are complete ($M_R = -14.0$). The slope of the best fit is consistent with the faint end of a Schechter luminosity function with $\alpha \sim -1.75 \pm 0.27$ (Schechter 1976), which is considerably steeper than that reported by Ribeiro, de Carvalho, and Zepf (1994) for the luminosity function of compact groups, $-0.82 \pm 0.15$. Marzke, Huchra, and Geller (1994) detect a similar excess of faint galaxies with respect to the bright end of the field galaxy luminosity function in the CfA Redshift Survey.

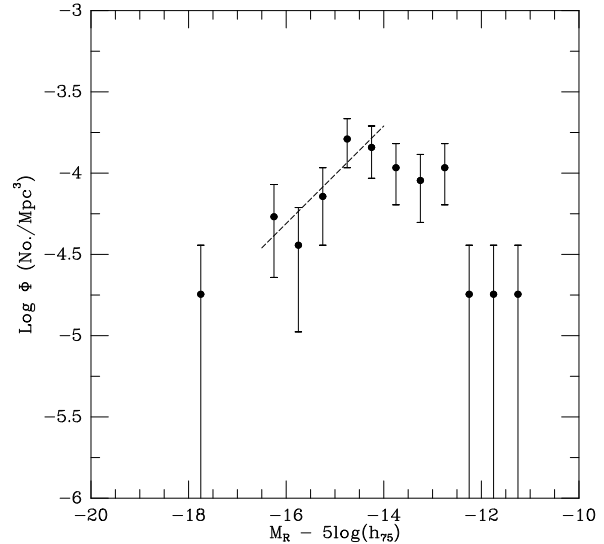

Fig. 2.— LUMINOSITY FUNCTION OF TIDAL DWARFS. The tidal dwarf magnitudes are binned in half-magnitude intervals and the error bars represent $1\sigma$ counting errors. The lack of objects fainter than $M_R = -14 + 5\log h_{75}$ reflects the incompleteness of our sample at these magnitudes. Fitting to the faint end of a Schechter luminosity function, the best slope (dashed line) is $\alpha \sim -1.75 \pm 0.27$.

Despite the reasonable agreement between our measurement of the faint end luminosity function and other previous results, we stress that our luminosity estimates, and in particular our luminosity function, are quite uncertain. First, some of the dwarf candidates may not be physically-bound systems or even in the compact group. Second, there are large uncertainties in the local sky background subtraction. Third, our R-band images include H$\alpha$ line emission and because these dwarfs may be actively forming stars, their luminosities may be contaminated by emission from ionized gas. However, at the typical distances of compact groups, $z = .03$, even a bright HII region will not be mistaken for a dwarf galaxy because its H$\alpha$ flux, $R \sim 24.4$, is well below the detection limit, $R \sim 21.5$. Despite the caveats mentioned above, we estimate the masses of these dwarf candidates, adopting a mass-to-light ratio of unity (consistent with the prediction of the Barnes and Hernquist 1992 model). The resulting histogram of dwarf masses is shown in Figure 3a. There is one exceptionally bright "dwarf" with an estimated mass of $6.37 \times 10^8 M_\odot$, which interestingly appears at the end of the tidal tail in HCG 001. The other tidal dwarfs have masses in the range $1.9 \times 10^6 M_\odot \leq M \leq 1.8 \times 10^8 M_\odot$ with a median value of $2.5 \times 10^7 M_\odot$. Diameters are estimated from the projected areas calculated by FOCAS (see Figure 3b). Our smallest candidate has an area of $\sim 13.5$ arcsec$^2$, so the candidates are well-resolved. Most diameters are between one and six $h_{75}^{-1}$ kpc with the median value $\sim 3 h_{75}^{-1}$ kpc. Table 3 summarizes the tidal dwarf data. Column 1 identifies individual tidal features, column 2 gives the distance of each tidal dwarf from the nucleus of the parent galaxy, and columns 3, 4, and 5 list magnitude, estimated mass, and diameter.

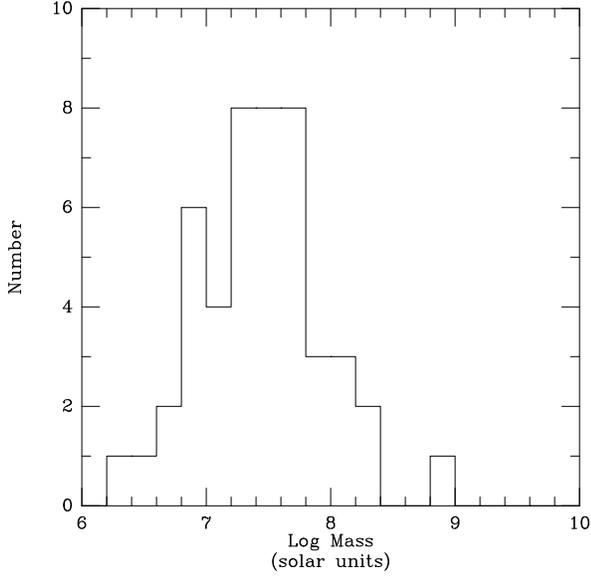

Fig. 3a.— MASS DISTRIBUTION OF TIDAL DWARFS. This distribution is based solely on luminosity and assumes M/L = 1. The median dwarf mass is $\sim 10^{7.4} M_\odot$.

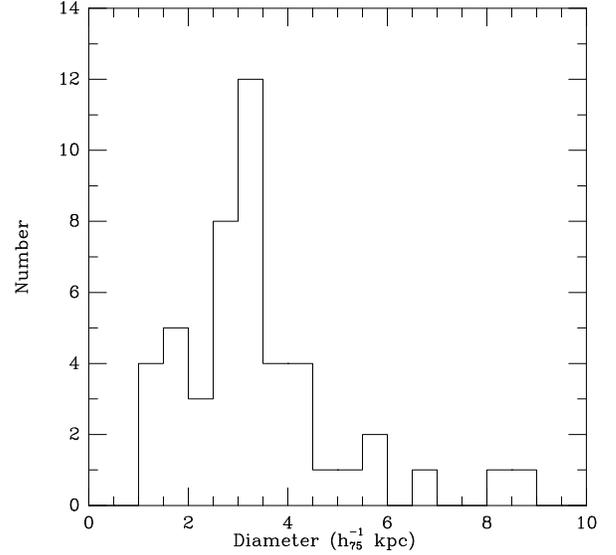

Fig. 3b.— SIZE DISTRIBUTION OF TIDAL DWARFS. This distribution is based on circularizing the projected area of objects detected by FOCAS. The median dwarf diameter is $\sim 3 h_{75}^{-1}$ kpc.

We now discuss the spatial and mass distribution of the dwarf candidates to provide general constraints on models of tidal dwarf formation. Two recent models, by Barnes and Hernquist (1992; hereafter BH) and Elmegreen et al. (1993; hereafter E93), provide a preliminary examination of the problem. While both models produce dwarf galaxies in tidal tails with the range of masses observed both in this study and those of isolated interacting galaxies (e.g., Mirabel et al. 1991, Mirabel et al. 1992, and Elmegreen et al. 1995), there is a basic difference in the mechanism through which the dwarfs form. In the BH model, stars in regions along the tail become gravitationally bound and the gas later falls into those potential wells. In the E93 model, regions of gas along the tail become Jeans unstable and collapse. Are the global characteristics of our candidates consistent with either model?

The magnitude distribution in each system is shown in Figure 4. Each row of points represents candidate tidal dwarfs belonging to a particular parent galaxy. In some cases (31a, 38b, 92b, and 92d) two tidal features are combined because they are associated with the same parent. Comparisons between systems are difficult with the current data because there are fewer than four tidal dwarfs in most systems and there are differences in the limiting magnitude between systems. We note two situations which may be indicative of a "preferred" magnitude range within a given system. First, there is a lack of objects with magnitudes brighter than $M_R - 5\log h_{75} = -13$ in system 16a. Second, we expect similar limiting magnitudes for systems 92b and 92c yet the faintest candidates have $M_R - 5\log h_{75} = -13.9$ and $-12.4$, respectively. A "preferred" magnitude range that varied among interactions, if confirmed with further observations, would support the E93 model of tidal dwarf formation.

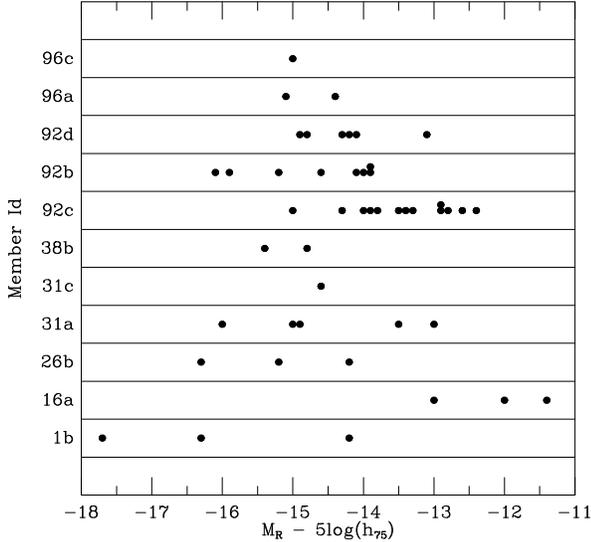

Fig. 4.— MAGNITUDE DISTRIBUTION PER PARENT GALAXY. The range of magnitudes for each parent galaxy is illustrated here. Is there a preferred magnitude range for each system or is the distribution of magnitudes completely random? The results are inconclusive because most systems have too few data points to perform a valid statistical analysis. Qualitatively, we do note a lack of bright objects in system 16a and a lack of faint objects in system 92b.

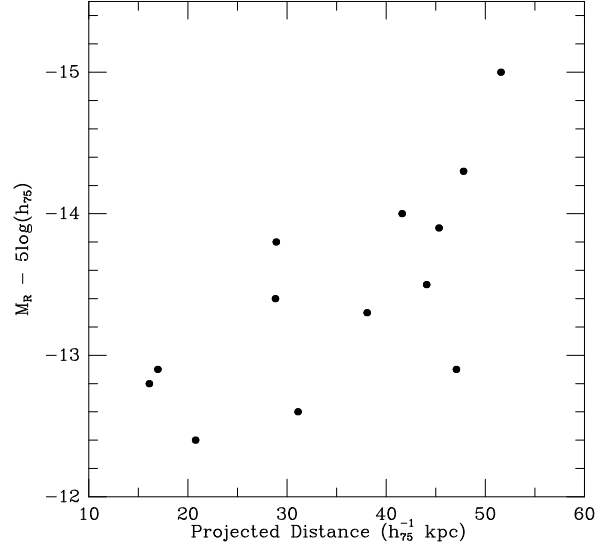

Fig. 5.— MAGNITUDE VS. PROJECTED DISTANCE FROM PARENT NUCLEUS (92C). With 99% confidence, there exists a linear correlation between magnitude and distance, i.e., the brightest dwarfs occur in the outer region of the tidal tail.

We now examine the distribution of dwarf masses along tidal features, beginning with the richest system, HCG 92c, to search for a correlation between magnitude and distance (see Figure 5). The data show a tendency for the dwarf luminosity to increase with distance from the parent. A Spearman rank correlation test gives a 99.3% probability that the distribution of magnitudes is not random. However, the HCG 92c tidal feature is exceptional in terms of size and richness. Because there is no apparent nearby perturber of HCG 92c, it is possible that the interacting nuclei have already merged. This tail may then be long-lived, which is thought to be atypical for the compact group environment. These data contradict the expectation that because only high mass objects survive the large tidal forces near the center of the parent galaxy (as seen in the BH simulations) there should be a lack of faint objects in the inner region of the tail. In fact, the brightest object in each tail/arm is located at the tip in five of the fifteen tidal features. This situation is analogous to that of the Antennae (Mirabel et al. 1992) and Superantennae (Mirabel et al. 1991). The presence of the brightest dwarf in the system at the end of the tidal feature is not predicted by the BH model, but can be accounted for by the E93 model if the parent has an extended gaseous disk. Clearly tidal disruption must play a role, although the inclusion of an extended gaseous disk in the parent galaxies appears to be most important.

Although both results cited above favor the E93 model, a definitive resolution of the tidal dwarf formation mechanism awaits detailed observations of the dwarfs themselves. Observations of their current star formation rates, their stellar populations, their metallicities, and the distribution of gas within them and within the tidal features are necessary.

## 4. Dwarf Formation in a Dense Environment

We now estimate the fraction of all compact group dwarfs formed in tidal debris by dividing the number of dwarfs we expect to have formed in tails by the total number of dwarfs in compact groups. The number of dwarfs in each group is determined by extrapolating the luminosity function of compact groups, which is measured to have a faint end slope of $\alpha \sim -0.82$ for a sample complete to $M_B = -15.1 + 5\log h_{75}$ (Ribeiro et al. 1994). There is some controversy regarding the slope of the compact group luminosity function because it is much flatter than that observed in clusters (Sandage, Binggeli, & Tammann 1985; Ferguson & Sandage 1991). If our adopted slope, $-0.82$, is too flat, we will underestimate the number of group dwarfs and overestimate the contribution of tidal dwarfs to the general population. We convert our $R$ magnitudes to $B$ by assuming a typical color for the tidal dwarfs of $(B-R) = 1.4$, the median value for clumps in the Superantennae (Mirabel et al. 1991). The magnitude interval of interest then corresponds to $-16.6 < M_B - 5\log h_{75} \leq -11.6$. By integrating the extrapolated luminosity function over this magnitude range we calculate a volume density of dwarf galaxies of $1.2 \times 10^{-4} h_{75}^3$ galaxies/Mpc$^3$. Dividing by the volume density of compact groups, $1.6 \times 10^{-5} h_{75}^3$ groups/Mpc$^3$ (Mendes de Oliveira & Hickson 1991), we obtain an estimate of approximately 7.5 dwarfs per group in our magnitude interval for $h_{75} = 1$.

The number of tidal dwarfs currently being produced in tidal tails in compact groups is estimated from our data. To convert the number of tidal dwarfs seen today to the total number of tidal dwarfs formed in compact groups we need to know the number of dwarfs formed in the past that are currently not seen in tidal features either because the tidal features and dwarfs have dispersed or because the dwarfs have merged with or been tidally disrupted by the parent galaxy. First, we estimate the average number of interactions among parent galaxies over the lifetime of a group. Let $\tau_g$ be the lifetime of a compact group and $\tau_t$ the lifetime of a tidal tail in this environment. If $f_t$ is the observed fraction of groups with tidal features, then the number of interactions, $i$, that produce tidal tails over $\tau_g$, can be expressed as $i = (\tau_g/\tau_t) \times f_t$.

The ratio of timescales $\tau_t/\tau_g$ for compact groups is uncertain, but can be estimated from simulations. In the merger of an isolated pair of galaxies (Barnes 1993) one can easily follow the development of tidal tails and unmistakable tidal features that exist for a significant part of the merger process. In a multiple merger (Barnes 1993) the resultant tidal debris is quickly scattered around the interacting galaxies. Tidal tails do sometimes form in a multi-galaxy environment when the interaction involves only two group members, but these tails are short-lived ($< \frac{1}{10}$ of the merger time) because they are disrupted by other galaxies. It is also difficult to distinguish such tidal features amidst the general debris from several simultaneous interactions. From these simulations, we estimate that $(\tau_t/\tau_g) \sim .10$. The number of interactions per group that produce tidal tails during its lifetime is then $i = (\tau_g/\tau_t) \times f_t = (1/.1) \times (7/42) = 1.7$. Within the eight observed interactions (HCG 092 has two interactions) there are 40 tidal dwarf candidates in the magnitude interval $-18 < M_R - 5\log h_{75} \leq -13$. This yields a production rate of 5.0 dwarfs per interaction.

The survival rate of tidal dwarfs must also be considered. Elmegreen et al. (1993) predict that the ultimate fate of any dwarf is linked to the mass ratio of the interacting galaxies. If the perturber mass is greater than the parent mass, the clumps in the tail will be ejected into the group at large or become satellites of the new galaxy. In the Elmegreen et al. (1993) simulations, tidal tails are produced regardless of the mass ratio, so we expect that the perturber mass is greater than the parent mass approximately half of the time and assume that half of the tidal dwarfs currently observed in the tidal tails will survive.

The estimated number of tidal dwarfs produced per group is (interactions per group) × (dwarfs per interaction) × (survival rate) = $1.7 \times 5.0 \times .5 = 4.2$. This is 56% of the expected number of dwarfs throughout an entire group (7.5 dwarfs per group). Because half of the tidal dwarfs identified are in HCG 092, the estimate of the fraction of tidal dwarfs is considerably lower ($\sim 30\%$) if this group is excluded from the analysis. A steeper compact group luminosity function or contamination of tidal dwarf candidates by background galaxies, foreground stars, and group dwarfs would also decrease the estimated tidal dwarf fraction. On the other hand, if some of the systems in our sample are not compact groups, then the fraction of tidal dwarfs will be underestimated because the number of interactions per real group would increase. Lastly, our estimate is conservative because our sample is probably not complete to $M_R = -13 + 5\log h_{75}$. Despite the large uncertainties, we conclude that the tidal dwarf formation mechanism is a significant contributor to the dwarf population of compact groups.

## 5. Summary

We conduct a systematic survey of tidal debris in the environment of Hickson compact groups of galaxies to address the question of whether a significant fraction of dwarf galaxies in dense environments form as a result of galaxy-galaxy interactions. In R-band images of 42 compact groups, we identify seven groups with tidal arms or tails. Within the tidal features, we find an average of 3.1 dwarf galaxy candidates (as few as 1 and as many as 13) per feature with magnitudes in the range $-17.7 \leq M_R - 5\log h_{75} \leq -11.4$ and a median magnitude of $M_R = -14.2 + 5\log h_{75}$. The resultant luminosity function for tidal dwarfs can be fit by a Schechter function with a faint end slope of $\alpha \sim -1.75 \pm 0.27$. This is much steeper than the reported compact group luminosity function (Ribeiro et al. 1994) and so tidal dwarfs may represent an excess faint population.

We find no general correlation between the mass of tidal clumps and their projected distances along the tails. However, the richest tidal tail in our sample (92c) shows a significant correlation with luminosity increasing with radius, and in five of fifteen tidal features the most luminous dwarf is at the end of the tidal feature. These results support the Elmegreen et al. (1993) models, which produce a large gas deposit at the end of the tidal tail opposite the perturbing galaxy.

We estimate the contribution of tidal dwarf formation and suggest that a substantial fraction (at least one-third and perhaps more than one-half) of all dwarfs in compact groups are formed during galaxy-galaxy interactions. Because tidal tails are observed in all environments, this formation mechanism could have more general implications. Confirmation of the candidates as tidal dwarfs will require $H\alpha$, colors, and spectroscopy to determine star formation rates, stellar populations, and if they are gravitationally bound members of the tidal feature. Our estimate of the importance of the tidal dwarf formation mechanism will also be improved by increasing the sample, by establishing the compact group luminosity function, and by conducting simulations that more reliably predict the relevant timescales and products. *We conclude that some galaxies do not form as described by the standard hierarchical models of galaxy formation and that the fraction of compact group dwarfs produced within tidal debris is not negligible.*

We would like to thank R. Ciardullo, L. Hoffman, M. Kaufman, V. Rubin, and E. Salpeter for all their helpful comments and Bob Hill and Skip Staples for their assistance at the telescope. DZ acknowledges financial support from the California Space Institute.

TABLE 2
SUMMARY OF TIDAL FEATURES

| HCG No. | Interacting Members | Tidal Dwarfs | Feature Description | MdeO/H Results |
|---|---|---|---|---|
| 01 | a-Sc, b-Im | 3 | b-tidal tail extending west, then north | a-tidal arm (I) <br> b-tidal arm (I) |
| 16 | a-SBab, b-Sab | 3 | a-tidal tail extending east, undetected by MdOH | a-disturbed, IR, lopsided r.c., radio (I) <br> b-tidal tails, peculiar r.c. (I) |
| 26 | a-Scd, b-E0 | 3 | b-tidal tail extending NW | a-asymmetric, IR, weak radio (I) <br> b-tail-like features, weak radio (I) |
| 31 | a-Sdm, c-Im | 6 | 2 tidal arms extending NE from both nuclei, 3rd tidal arm extending SE | a-disturbed, possible IR, sinusoidal r.c., WR (I,M) <br> c-peculiar morph., IR, peculiar r.c. (I,M) |
| 38 | b-SBd, c-Im | 2 | b-tidal arms north and south of nucleus | b-tidal tail, disturbed, IR (I,M) <br> c-disturbed, possible IR (I,M) |
| 92 | c-SBa, ? | 13 | c-tidal tail extending SE | c-tidal arm, Seyfert, IR, radio (I,M) |
| 92 | b-Sbc, d-SB0 | 14 | tidal tails north, undetected by MdOH; tidal arms south | b-tidal arm (I) <br> d-tidal arm, weak radio (I,M) |
| 96 | a-Sc, c-Sa | 3 | a-tidal tail extending NW <br> c-tidal tail extending east, undetected by MdOH | a-tail-like feature, Seyfert, IR, strong radio (I,M) <br> c-extension along axis, poss. IR, radio (I) |

NOTE.—

| | | | | |
|---|---|---|---|---|
| r.c. | rotation curve | | (I) | interacting |
| IR | infrared | | (M) | merging |
| WR | Wolf-Rayet spectral features | | | |

TABLE 3
Tidal Dwarf Summary

| Name | Distance ($h_{75}^{-1}$kpc) | Magnitude ($M_R - 5\log h_{75}$) | Mass [$\log(\frac{M}{M_\odot})$] | Diameter ($h_{75}^{-1}$kpc) |
|---|---|---|---|---|
| 01b | 13.66 | −16.3 | 8.2 | 8.98 |
|  | 16.72 | −14.2 | 7.4 | 3.33 |
|  | 20.82 | −17.7* | 8.8 | 8.26 |
| 16a | 12.40 | −13.0 | 6.9 | 1.89 |
|  | 17.28 | −12.0 | 6.5 | 2.00 |
|  | 26.62 | −11.4 | 6.3 | 1.14 |
| 26b | 22.71 | −16.3 | 8.2 | 6.72 |
|  | 27.03 | −15.2 | 7.8 | 5.57 |
|  | 29.36 | −14.2 | 7.4 | 5.75 |
| 31a N | 5.25 | −15.0 | 7.7 | 2.72 |
|  | 7.16 | −14.9 | 7.7 | 2.46 |
|  | 9.22 | −13.5 | 7.1 | 1.31 |
| 31c N | 2.15 | −14.6 | 7.6 | 1.26 |
| 31a S | 5.04 | −13.0 | 6.9 | 1.18 |
|  | 6.94 | −16.0* | 8.1 | 2.59 |
| 38b N | 6.29 | −14.8 | 7.6 | 4.15 |
| 38b S | 8.10 | −15.4 | 7.9 | 2.99 |
| 92c | 16.10 | −12.8 | 6.8 | 3.32 |
|  | 16.97 | −12.9 | 6.9 | 3.00 |
|  | 20.78 | −12.4 | 6.7 | 2.01 |
|  | 28.82 | −13.4 | 7.1 | 3.89 |
|  | 28.93 | −13.8 | 7.2 | 2.85 |
|  | 31.11 | −12.6 | 6.8 | 1.83 |
|  | 38.08 | −13.3 | 7.0 | 3.26 |
|  | 41.60 | −14.0 | 7.3 | 3.15 |
|  | 44.09 | −13.5 | 7.1 | 3.61 |
|  | 45.33 | −13.9 | 7.3 | 3.00 |
|  | 47.08 | −12.9 | 6.9 | 3.16 |
|  | 47.81 | −14.3 | 7.4 | 4.13 |
|  | 51.57 | −15.0* | 7.7 | 4.21 |
| 92b S | 9.18 | −14.6 | 7.6 | 2.81 |
|  | 9.89 | −14.0 | 7.3 | 1.95 |
|  | 10.12 | −15.2 | 7.8 | 3.04 |
|  | 10.30 | −15.9 | 8.1 | 3.97 |
|  | 10.81 | −13.9 | 7.3 | 2.63 |
| 92b N | 13.25 | −13.9 | 7.3 | 3.48 |
|  | 21.64 | −16.1 | 8.2 | 3.74 |
|  | 26.52 | −14.1 | 7.4 | 2.84 |
| 92d S | 7.47 | −14.9 | 7.7 | 2.92 |
|  | 8.93 | −14.1 | 7.4 | 1.92 |
| 92d N | 21.03 | −14.2 | 7.4 | 3.49 |
|  | 22.83 | −13.1 | 7.0 | 1.74 |
|  | 24.97 | −14.8 | 7.6 | 3.10 |
|  | 31.14 | −14.3 | 7.4 | 3.08 |
| 96c E | 62.15 | −15.0* | 7.7 | 4.72 |
| 96a W | 27.39 | −14.4 | 7.5 | 5.09 |
|  | 39.71 | −15.1* | 7.8 | 4.41 |

NOTE.— An asterisk following the magnitude indicates a bright dwarf candidate located at the end of a tidal tail.

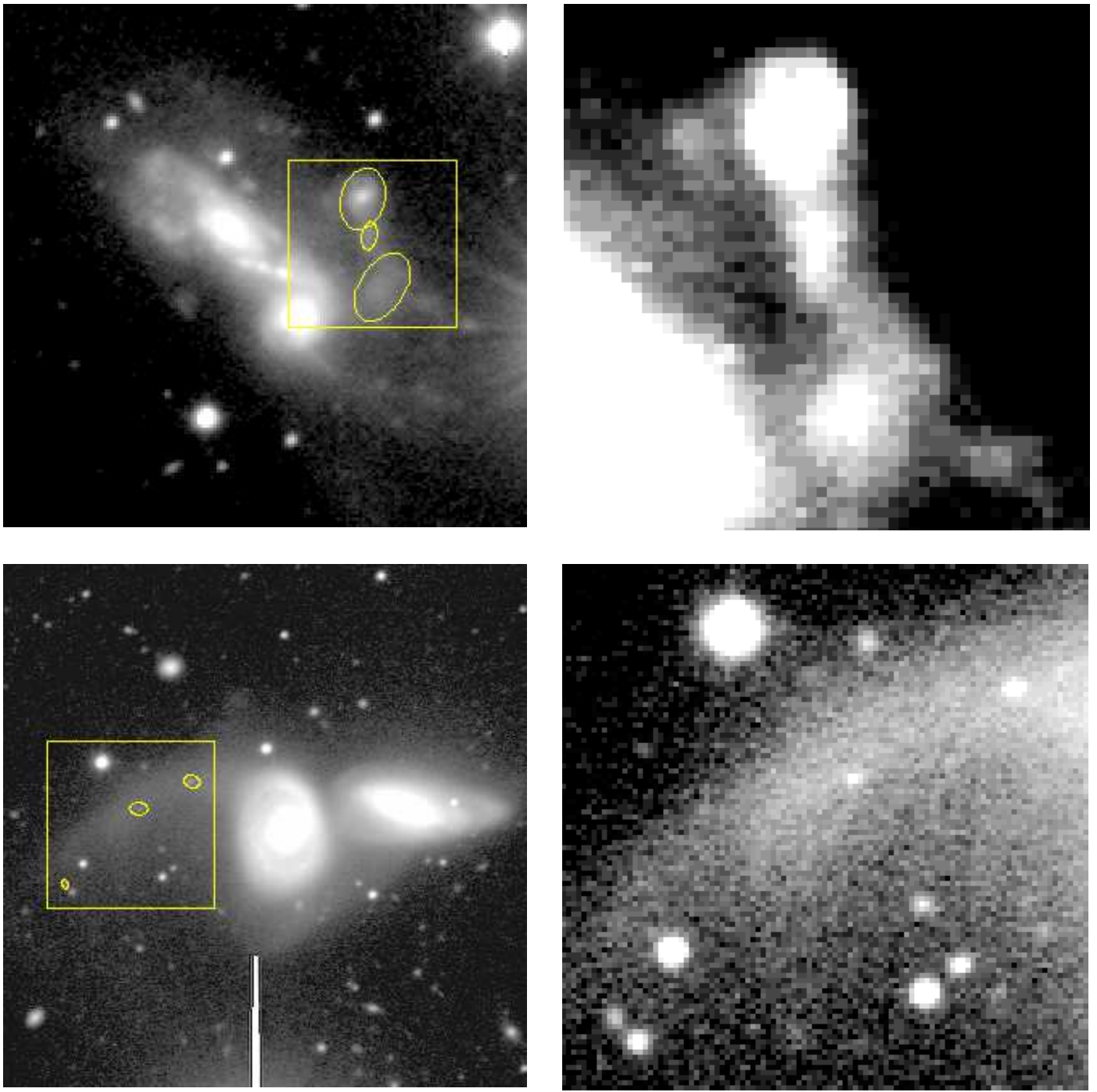

Fig. 1.— R-band images of interacting galaxies in our sample of Hickson compact groups. The first three panels contain full images on the left and expanded and rescaled images of the dwarf candidates on the right. The FOCAS-detected "tidal dwarf candidates" are encircled. Full frame sizes are $2' \times 2'$ for all groups except 16 and 96 which are $4' \times 4'$. Figure 1a contains images of HCG 001 (top) and 016 (bottom).

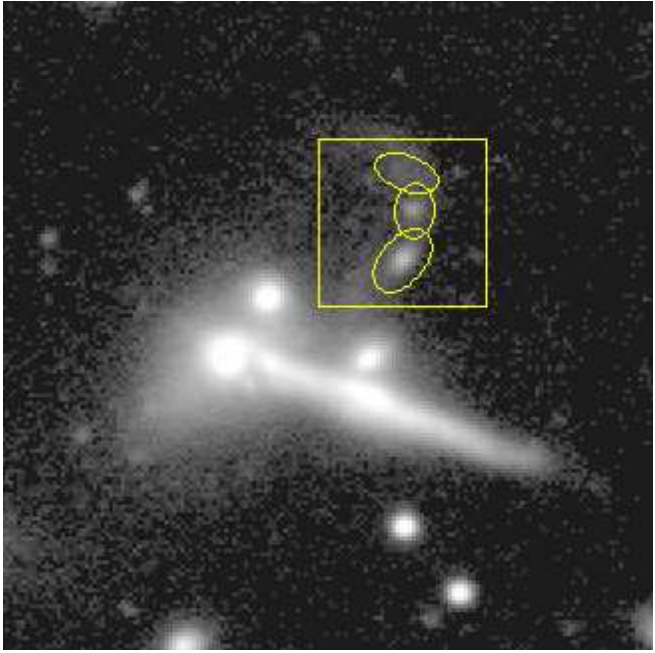
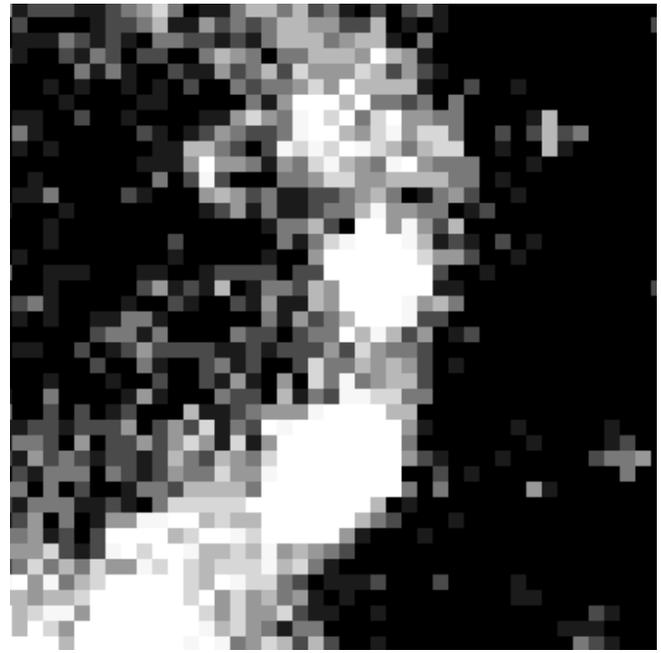
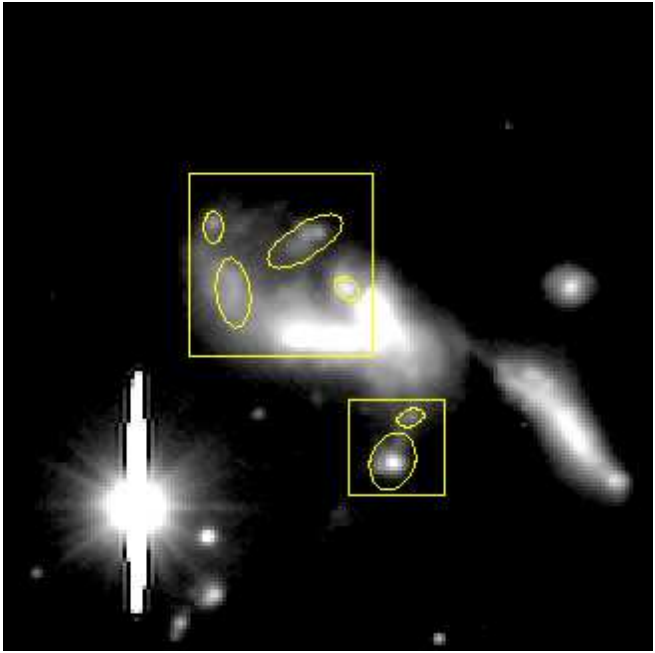
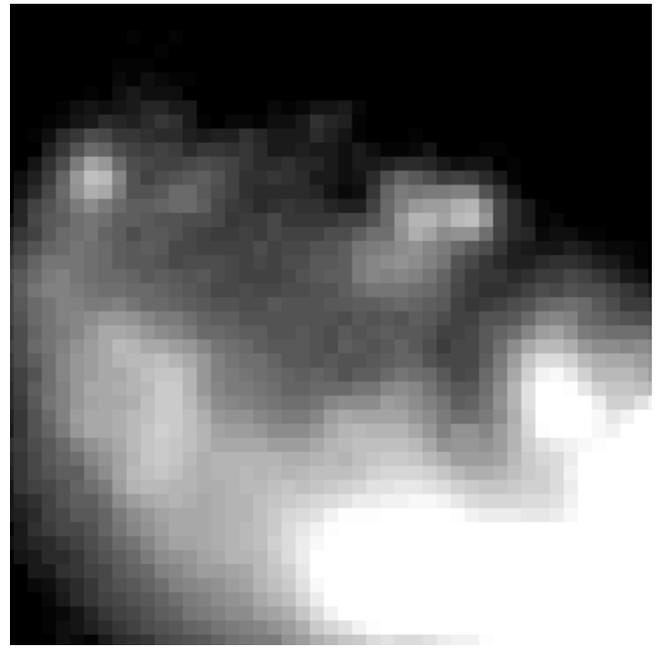
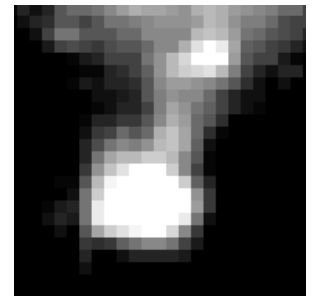

Fig. 1b.— IMAGES OF HCG 026 (top) AND 031 (bottom).

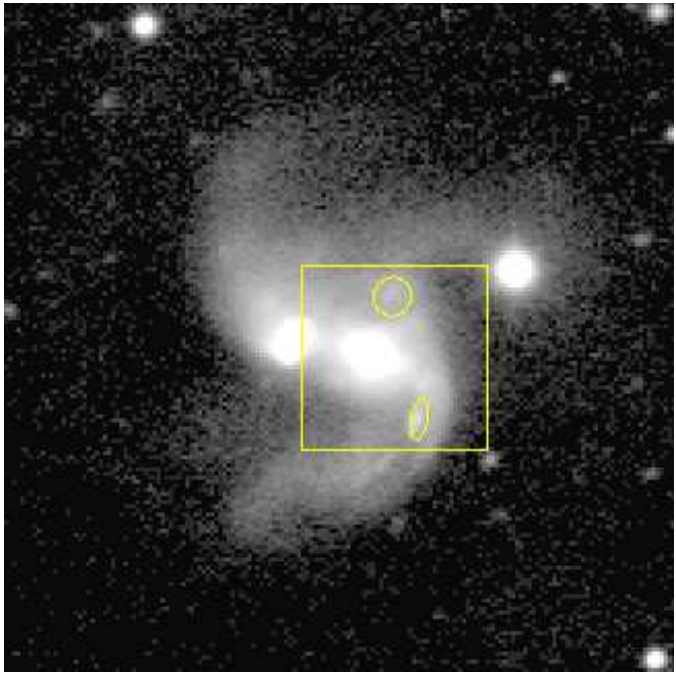
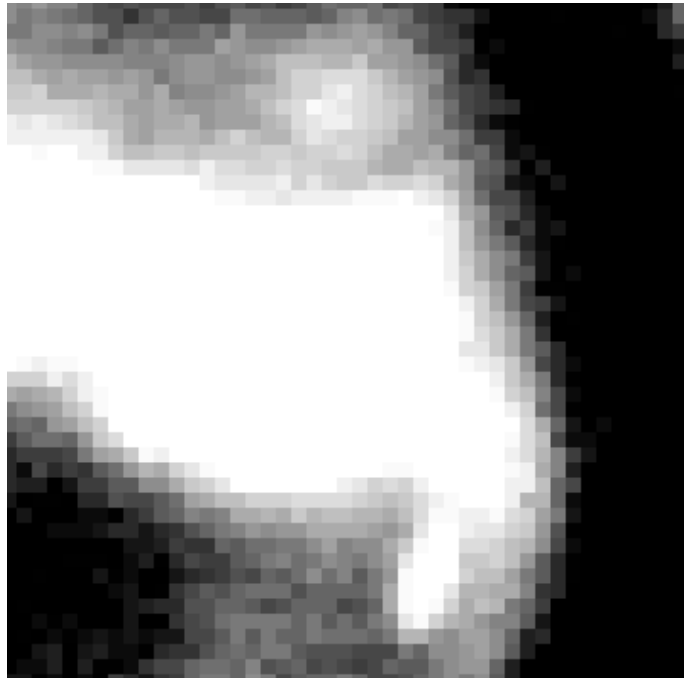
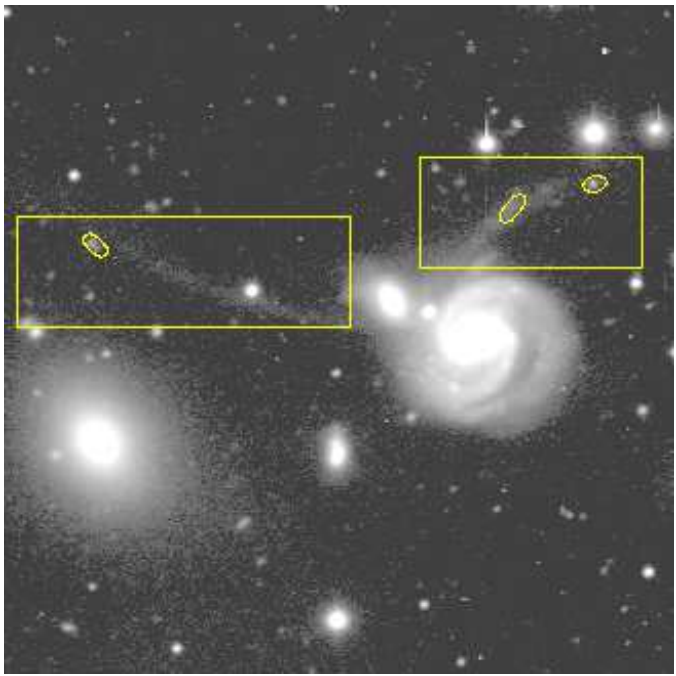
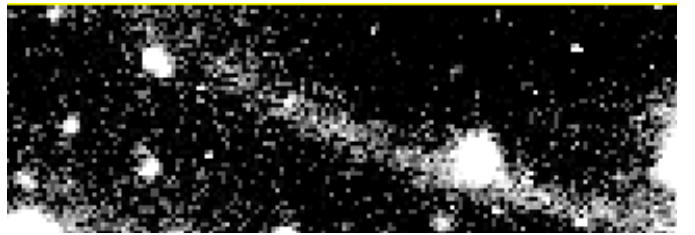
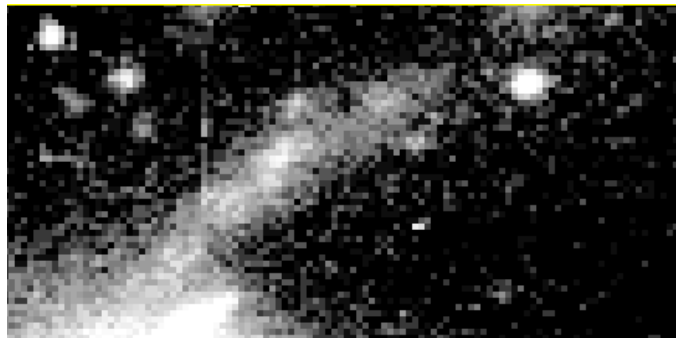

Fig. 1c.— IMAGES OF HCG 038 (top) AND 096 (bottom).

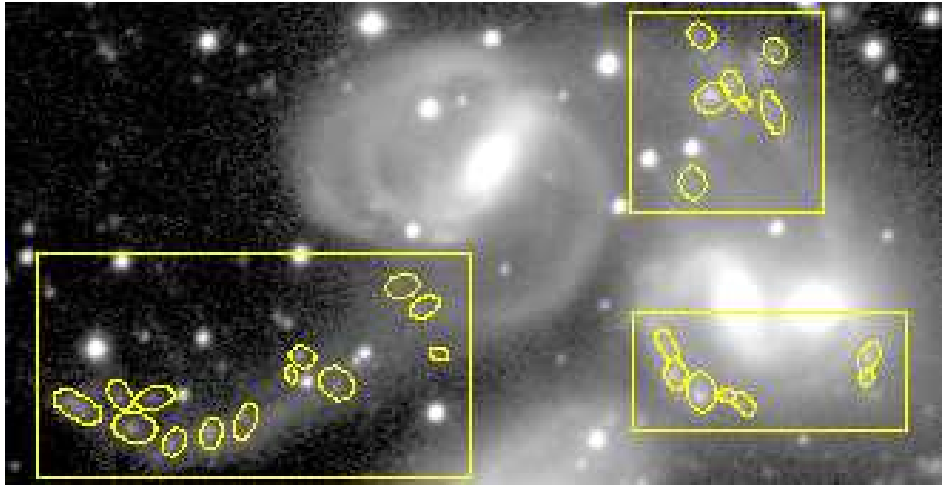

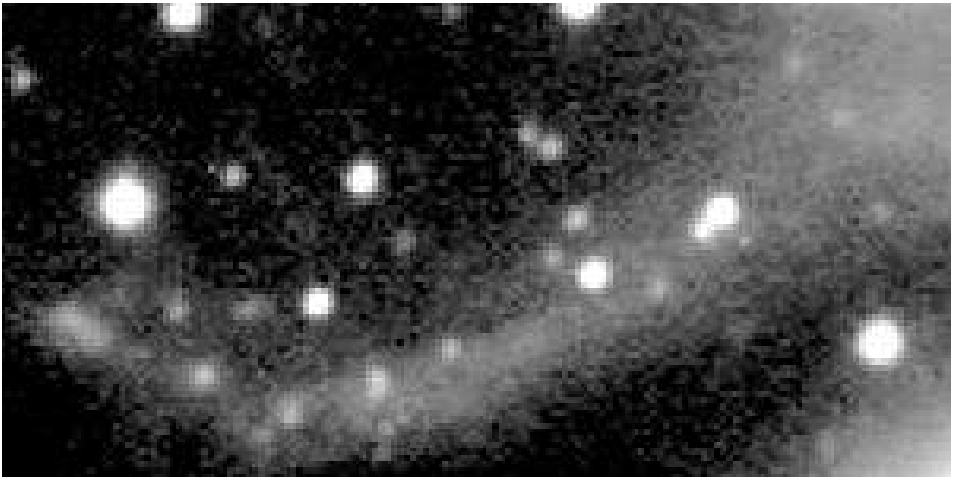

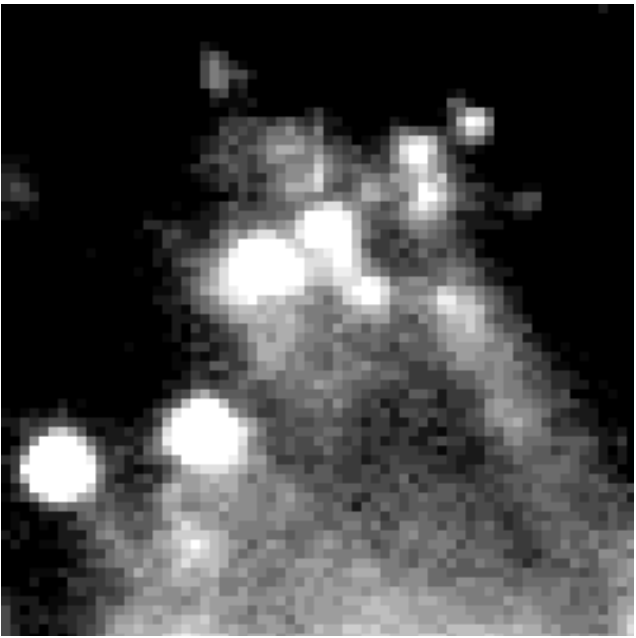

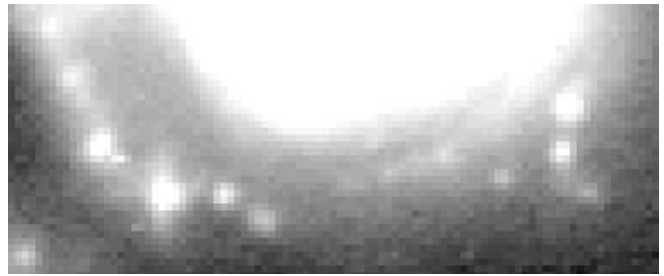

Fig. 1d— IMAGES OF HCG 092. At the top of the page is a group image. Member galaxy 92c appears to the left in the image and two interacting galaxies, 92b and 92d, are on the right. The width of the frame is $\sim 6'$. In the middle of the page is an enlargement of the tidal tail in 092c. At the bottom are the regions above and below the interacting pair 092bd.